# Large publishing consortia produce higher citation impact research but co-author contributions are hard to evaluate[1]

Mike Thelwall, Statistical Cybermetrics Research Group, University of Wolverhampton, UK.

This paper introduces a simple agglomerative clustering method to identify large publishing consortia with at least 20 authors and 80% shared authorship between articles. Based on Scopus journal articles 1996-2018, under these criteria, nearly all (88%) of the large consortia published research with citation impact above the world average, with the exceptions being mainly the newer consortia for which average citation counts are unreliable. On average, consortium research had almost double (1.95) the world average citation impact on the log scale used (Mean Normalised Log Citation Score). At least partial alphabetical author ordering was the norm in most consortia. The 250 largest consortia were for nuclear physics and astronomy around expensive equipment, and for predominantly health-related issues in genomics, medicine, public health, microbiology and neuropsychology. For the health-related issues, except for the first and last few authors, authorship seem to primary indicate contributions to the shared project infrastructure necessary to gather the raw data. It is impossible for research evaluators to identify the contributions of individual authors in the huge alphabetical consortia of physics and astronomy, and problematic for the middle and end authors of health-related consortia. For small scale evaluations, authorship contribution statements could be used, when available.
**Keywords**: Scientometrics; Collaboration; Research consortia; Research impact; Citation impact; MNLCS

## 1   Introduction

The frequency and scale of academic research collaboration varies within and between fields but has grown steadily over time (Fortunato, Bergstrom, Börner, Evans, Helbing, et al., 2018). Whilst arts and humanities scholars often work alone (Larivière, Gingras, & Archambault, 2006; Wuchty, Jones, & Uzzi, 2007), some types of experimental and applied research require large teams (de Solla Price, 1986). Interdisciplinary collaboration may be needed to solve complex real-world problems (e.g., Moore, Moore, Billiar, Vodovotz, Banerjee, & Moldawer, 2017; Olds, 2016), in a process that has been called mode 2 science (Gibbons, Limoges, Nowotny, Schwartzman, Scott, & Trow, 1994). Big teams may also be needed to operate large scale equipment, such as the hadron collider, and collaboration may sometimes be more efficient by combining expertise. Weak or strong collaborations have also been formed to create specific resources of common value, such as the Human Genome Project (Roberts, 2001) leading to their landmark paper (The International Human Genome Sequencing Consortium, 2001), epigenomics mapping (Bernstein, Stamatoyannopoulos, Costello, Ren, Milosavljevic, et al., 2010) and gene product function ontologies (Gene Ontology Consortium, 2014). Nevertheless, research consortia may be formed through policy decisions or in response to research funding requirements, which may be less successful (Defazio, Lockett, & Wright, 2009) or primarily benefit less successful partners (Hoekman, Scherngell, Frenken, & Tijssen, 2013).

Close collaboration can also occur around data when a discipline collaborates to create standards and repositories to systematically exchange information (e.g., AACR Project GENIE Consortium, 2017; Buniello, MacArthur, Cerezo, Harris, Hayhurst, Malangone, & Suveges, 2018; Costa, Qin, & Bratt, 2016; Vermeulen, Parker, & Penders, 2013; Welter, MacArthur, Morales, Burdett, Hall, Junkins, & Parkinson, 2013). These are problematic to evaluate with scientometric methods because the primary output of such collaborations is not a set of journal articles.

A more recent reason to collaborate on some types of research is that competing groups using similar methods to address a shared problem may lead to a situation in which the wrong answer achieves statistical significance and is published, whereas combining the competing teams would give the correct result (Ioannidis, 2005; Munafò, Nosek, Bishop, Button, Chambers, Du Sert, & Ioannidis, 2017). This occurred for Genome-Wide Association Studies (GWAS), for example. In this field, multiple teams attempted to identify single nucleotide polymorphisms (SNPs) on the human genome that associated with genetic diseases or health problems. Each team generated its own dataset, typically consisting of hundreds or thousands of human volunteers and genetic information on a sample of SNPs. Since only positive results were published and many negative results were discarded, this led to incorrect results being published when they accidentally achieved statistical significance, unless more stringent thresholds were used (Pe'er, Yelensky, Altshuler, & Daly, 2008). A side-effect was the "winner's curse", with the effect of SNPs being overestimated in studies that found them (Garner, 2007). The solution adopted by the community included pooling datasets to improve statistical power so that the number of discarded negative results would be reduced. In this situation the incentive to form consortia was statistical and improved research validity and power (Psaty, O'Donnell, Gudnason, Lunetta, Folsom, Rotter, & Boerwinkle, 2009).

Collaboratively authored journal articles may be valued by their co-authors in research evaluation exercises because they tend to be more cited than solo articles in many fields, although the citation advantage of collaboration has diminished over time (Larivière, Gingras, Sugimoto, & Tsou, 2015). Although an advantage of team science is the ability to apply novel combinations of expertise, the research produced is not more novel than less collaborative research in some fields (Wagner, Whetsell, & Mukherjee, in press). The GWAS example above illustrates that large teams may be needed to make research more predictable. Author ordering may be a problem in large collaborations, making it difficult to assess individual scholars' contributions. Although alphabetical author ordering is not the norm in any broad field of science (Levitt & Thelwall, 2013; Waltman, 2012), it seems unlikely that a large group of authors on a long-term project would be able to precisely determine relative contributions and partial or full alphabetical ordering would be a logical solution to this. A common biomedical solution is to order the main contributors at the start and end of the author list non-alphabetically, with the remaining authors ordered alphabetically in the middle (Mongeon, Smith, Joyal, & Larivière, 2017).

Despite much evidence about collaboration in general within fields, science wide, or from an individual funding programme, no previous science-wide study has investigated the prevalence of publishing consortiums, their impact or use of alphabetical ordering. This is important because research evaluation exercises often need to evaluate the contributions of individual authors or departments, which may be difficult for large collaborative papers. This article uses a heuristic to identify publishing consortia throughout science, 1996-2018 to investigate impact and authorship order, and a manual classification of their type for deeper insights into their contributions to science.

- RQ1: Do large publishing consortia produce higher impact research?
- RQ2: Does authorship order reflect contributions in large publishing consortia?
- RQ3: Which types (fields, reason for existence) of large publishing consortiums exist?

## 2   Methods

All 67,608,187 Scopus records of type journal article published between 1996 and 2018 were downloaded between November 2018 and February 2019. Articles tagged with other categories, such as review papers or editorials, were excluded to focus on a standard document type. The data included the author names, as recorded by Scopus, up to a maximum (usually) of 100. Each author was also assigned a unique ID by Scopus and these IDs were used to compare authors even though they are likely to be sometimes incorrect to reduce errors caused by authors having the same common name. Scopus author IDs are automatically assigned by an author name disambiguation algorithm that prioritises precision (i.e., attempts to avoid conflating different researchers with the same name) (Moed & Plume, 2013). It accuracy seems to be high since a study of Japanese funded authors found it to have precision of 99% (99% of papers associated with an author ID were correctly assigned) and recall of 98% (98% of an author's papers were assigned to their ID) (Kawashima & Tomizawa, 2015). Duplicate IDs occurred in these lists when authors had multiple affiliations, reducing the total number of unique authors, for articles with large sets. For example, a paper with thousands of authors would have 100 author IDs in Scopus, with as few as 30 being unique if they had about three affiliations each. Duplicate IDs were removed before subsequent processing.

Document clustering has previously been used to identify scientific topics (Sjögårde & Ahlgren, 2018; Waltman, Van Eck, & Noyons, 2010) and network mapping has been used to identify collaboration patterns (Xie, Ouyang, & Li, 2016), but the task here is different: to identify groups of articles created by a mostly common set of authors. A heuristic was used to identify publishing consortia. A set of articles was classified as a large consortium product when all had at least 20 (unique) authors, each article had at least 80% of authors in common with one other article in the set, and the collection contained at least 3 articles, building collections through a simple agglomerative clustering algorithm. These values (80%, 20 and, less critically, 3) were chosen after manual inspection of the data with the aim of getting the most generous definition that included few false consortia (i.e. one that did not self-define as a coherent group). A commonality of 80% allows for moderate changes in the composition of a consortium.

The consortium identification method used would be defeated by consortia with hundreds of authors that substantially changed author name orders, since the number of authors in common would then be hidden by the Scopus 100 author maximum threshold. No examples were found of consortia that were split into separate groups because of this but others may not have been found. The single linkage agglomeration method allows a consortium to gradually evolve. For example, a group of authors that maintained a size of over 20 people, published at least one article every year, and had a personnel changeover rate of under 20% would appear as a single cluster of papers. This is because an article with, say, 20 authors would cluster with a second if up to four of these authors changed. It would cluster with a third paper with an overlap of less than 80% if another paper in the cluster had an overlap of at least 80% (Figure 1).

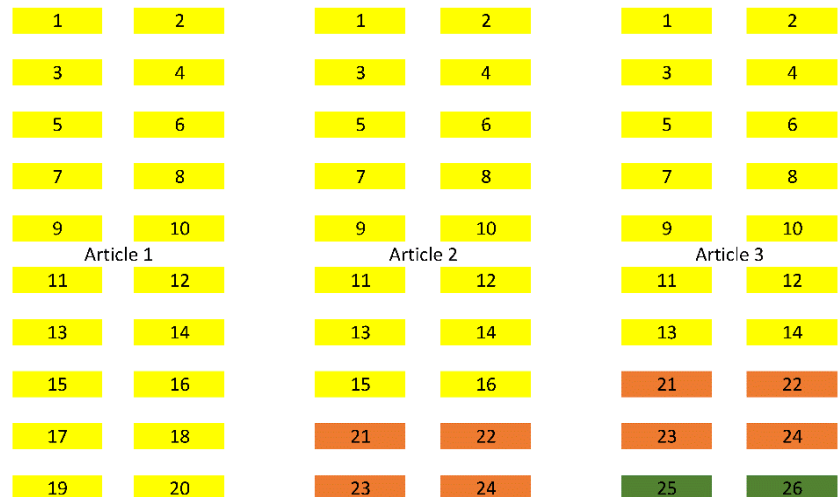

Figure 1. Three articles with partially overlapping authors clustered into one group. Colours indicate authorship overlaps. Although articles 1 and 3 have fewer than 80% authors in common, they both have 80% in common with article 2 and so are grouped into a single cluster.

In partial validation of the method and parameters chosen, most of the consortia identified had given themselves names and so were formal research groupings. The clustering rules may still have been too conservative, however, since there is no evidence about the extent to which clusters were missed. In particular, research groupings with selective authorship (e.g., only contributing authors from a consortium pool) would be omitted.

RQ1: The Mean Normalised Log Citation Score (MNLCS) (Thelwall, 2017) was calculated for each consortium to assess whether it produced research with a citation impact above the world average. This is a field normalised citation impact indicator that divides the citation score for each paper by the world average of all papers published in the same field and year. This avoids advantages for older articles (because they are normalised against contemporary articles) and for high citation fields (because they are normalised against articles from the same field). The log transformation ln(1+c) was applied to all citation counts before any calculation to reduces the impact of individual highly cited articles so that (a) the data is less skewed and it is more reasonable to average it with the arithmetic mean and (b) the resulting figure is less subject to variability (Thelwall & Fairclough, 2017). Most articles were in multiple fields and in these cases the arithmetic mean of the values for each field was used.

RQ2: Alphabetical ordering of authors is not straightforward to assess since a consortium may use partial or full alphabetical ordering (Mongeon, Smith, Joyal, & Larivière, 2017) and authors may fall into alphabetical order by accident. Partial alphabetical ordering can also be a side-effect of cultural differences in contributions since, for example, Chinese last names are more likely to start with Z than American last names. A non-alphabetic systematic ordering may also be used for articles originally written in languages, such as Chinese, that do not use the Latin alphabet (but alphabetical ordering from China is rare: Liu & Fang, 2014), and may use radical and stroke sorting or pronunciation order. Moreover, consortia may make different decisions about how to deal with issues such as prefixes (and their capitalisation), hyphenated names, double last names, accented characters in the extended Latin alphabet, cultures where the second name is the given name, cultures where the penultimate name is the family name, and whether to use the full first name(s) or the initials when ordering. Because of these issues, articles in perfect alphabetical (or equivalent) order according to the rules of a consortium may appear to be not lexically sorted based on the information available to Scopus or in the original article, in addition to any transcription errors. A heuristic was therefore used to judge the degree of alphabetical ordering. This was designed to be sensitive to partial alphabetical orderings after observations of this in some consortia. The alphabetical ordering heuristic used was to count the proportion of consecutive distinct (with different last names and/or first initials) authors that were in alphabetical order, according to their last name and first initial, as recorded in Scopus. Thus a score of 18/19 for an article with 20 authors indicates that then *n*th author was alphabetically before the *(n+1)*th author for all consecutive pairs of distinct authors except one. On average, an article should score 50% and values that substantially deviate from this indicate a degree of deliberate alphabetical order. For convenience of analysis, articles were split into groups based on the extent of alphabetical author ordering: Close to alphabetical (90%+); Partial alphabetical (between 60% and 99%); Close to non-alphabetical (from 40% to 60%); and Anti-alphabetic (Below 40%). A large team biomedical article with an average sized set (60%) of alphabetically ordered middle authors and half of the remaining authors accidentally alphabetically ordered by accident (Mongeon, Smith, Joyal, & Larivière, 2017) should therefore score 80% and fall comfortably into the partial alphabetical set. The Scopus limit to 100 author affiliations affects the accuracy of this, for a minority of the consortia. Although the average number of authors for a consortium article was 38, consortia with over 100 and partial alphabetical ordering might have this hidden if most of the first 100 authors were in an initial non-alphabetical section, although this seems unlikely.

RQ3: The largest 250 consortia were examined (all those with at least 17 papers) and manually clustered by discipline and consortium type. This clustering was performed by examining basic properties of the consortium and, from acknowledgements, methods descriptions and web searches, identifying the scope of the grouping. This was an inductive and iterative process, with the initial classifications being revisited and subsequently re-classified into a smaller number of different types. The largest consortium of each type was then singled out for a more detailed description, based on a brief publication history and, if relevant, web explorations. Physics and astronomy consortia appeared to be very similar and were ignored after the first few.

The key data behind the results, including a list of consortia, is on Figshare (https://doi.org/10.6084/m9.figshare.8214050.v1).

## 3   Results

Based on Scopus journal articles 1996-2018, the largest apparent consortium has 755 articles and 3927 consortiums have at least 3 articles, totalling 31,340 papers. This is 0.05% of the set examined. The largest 250 consortia include between 16 and 755 articles, accounting for 14,135 articles in total. Thus, large publishing consortia are numerically common in academia but are very rare overall, at least as defined by the heuristic used here.

## 3.1 Consortium size

Consortium size follows a power law, which gives a straight-line shape on a dual logarithmic scale (Figure 2). There are many small consortia (in terms of the number of papers published) and few large ones, with no evidence of any size being unusually common. A power law can be produced if there is a positive feedback mechanism but it is not clear how such a mechanism could operate here.

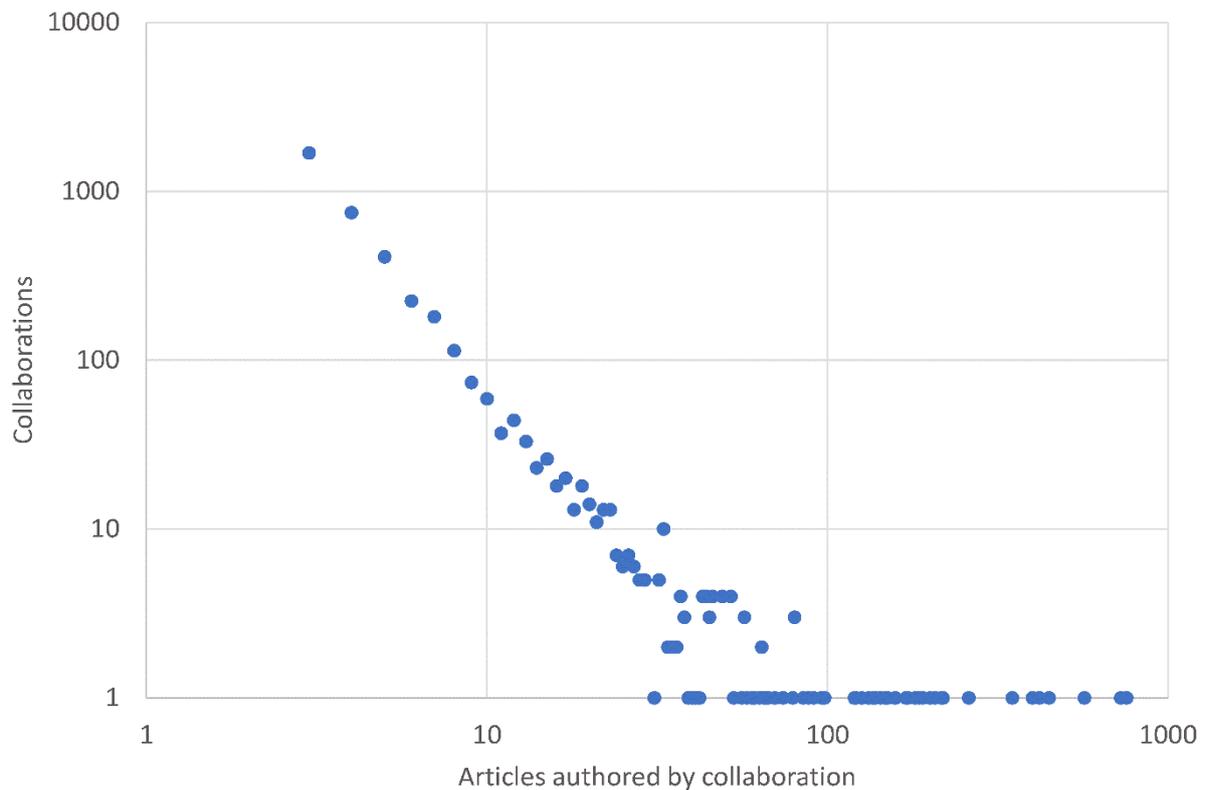

Figure 2. Sizes of consortia identified in Scopus publications 1996-2018.

## 3.2 RQ1: Consortium citation impact

The 3927 publishing consortia produced research with an average citation impact of MNLCS 1.954. On the log-normalised scale, this is almost double the impact of the world average (1). Because of the (typically) shrinking effect of the log normalisation part of MNLCS, large publishing consortia usually publish research with several times more citation impact than the world average, although the exact ratio depends on the average (untransformed) citation counts involved.

   The overwhelming majority (87.5%: 3427) of individual consortia had an MNLCS above the world average of 1, including the largest 93. The first MNLCS below 1 was for research from a single department (in the USA) rather than a multi-institution consortium. A Spearman correlation between latest year of the first published article from a consortium and MNLCS of -0.818 (p=0.000) confirms that consortia with a lower impact score tend to be the newer ones for which the MNLCS value is less reliable due to a shorter time to accrue citations. In fact, for very recently published articles, most have a normalised impact below the mean as an artefact of the dominance of zeros, further explaining this result (e.g., to give an extreme example, if only 1 out of 100 recently published articles had been cited then 99 of them would have a field and year normalised citation score of 0, which is below the world average of 1). All the 1,050 consortia with their first articles published before 2006 (for which citation counts are mature and hence most robust) had an average impact above the world average (i.e., MNLCS>1). This is clear statistical evidence that large publishing consortia reliably produce research with citation impact substantially above the world average.

A tiny positive Spearman correlation between cluster size (number of articles published by a consortium) and MNLCS of 0.072 (p=0.000) suggests that the number of articles published by a consortium has little relationship with the average impact of its research. This may be due to a variety of conflicting factors, however, such as consortium age, type and discipline.

## 3.3  RQ2: Author alphabetical order

Although two fifths (38%) of the consortia largely avoid any kind of alphabetical ordering (on average for their papers), a fifth (22%) are in perfect or close to perfect alphabetical ordering and two fifths (38%) are in partial alphabetical order (over 60% of consecutive different author names being in alphabetical order but less than 90%) (Table 1). Counting by paper rather than by consortium, three quarters of papers in consortia (75%) have at least partial alphabetical ordering (above 60%). This figure is higher than for consortia (60%) because the biggest consortia are more likely to use alphabetical ordering. The largest 19 consortia all have average alphabetical ordering rates of at least 84%, with 16 of them above 90%.

Table 1. Extent of alphabetical ordering, as judged by the proportion of consecutive pairs of distinct authors in the author list recorded in Scopus that were in alphabetical order, based on their last name and first initial. For reference, a long list of random authors would have an alphabetical score of 50%. For a consortium, the classification is based on the average extent of alphabetical ordering across all of its papers.

| Author order | Consortia | Papers |
|---|---|---|
| Close to alphabetical: At least 90% | 873 (22%) | 13622 (43%) |
| Partial alphabetical: Above 60% and below 90% | 1488 (38%) | 10028 (32%) |
| Close to non-alphabetical: From 40% to 60% | 1546 (39%) | 7623 (24%) |
| Anti-alphabetic: Below 40% | 20 (1%) | 67 (0%) |
| **Total** | **3927** | **31340** |

## 3.4  RQ3: Consortium types

Eight types of consortia were identified from manual examinations of the papers written by the largest 250 (Table 2). The types are distinguished based on the apparent consortium purpose and broad field. This typology is one coherent way of differentiating between consortia based on their apparent purpose, rather than a definitive typology. The largest example of each type is discussed below. Most of the types of consortia are organised around a specific narrow task or equipment, although some are subcollections within a department or funded broad research group.

Table 2. Unique combinations of fields and tasks for consortia publishing at least 17 papers, with examples of projects.

| Field | Type | Examples (no. of papers) |
|---|---|---|
| Nuclear Physics | A series of experiments or data analysis with equipment for measuring particles, such as at the Large Hadron Collider. | ATLAS Collaboration (755); CLEO Collaboration (400); D0 Collaboration (349) |
| Astronomy | Users of an expensive type of telescope (e.g., "High Energy Stereoscopic System" telescopes). | HESS Collaboration (150); MAGIC Collaboration (138); Fermi-LAT Collaboration (136) |
| Genomics | A shared genomics methods-based generic task (e.g., Genomic Encyclopedia of Bacteria and Archaea project with two research groups; and, "a semiautomated high-throughput pipeline at the Joint Center for Structural Genomics") or a disease-based task | GEBA project (148); Joint Center for Structural Genomics (85); Consortium on the Genetics of Schizophrenia (27); OCD Collaborative Genetics Study (25); IMAGE project (21); Cancer Genome Atlas Research Network (19) |
| Medicine | Registry or longitudinal cohort study for disease, based upon large collaboration at multiple sites (e.g., "Swiss [] outpatient clinics [] HIV patients.") | Swiss HIV Cohort Study (120); CONFIRM registry (52); Systemic Lupus [] cohort (32); US Extrahepatic Biliary Malignancy Consortium (26); Canadian Scleroderma Research Group (20); Chronic Total Occlusion Multicenter US Registry (19); IDACO Investigators (19); EUROCARE Working Group (18) |
| Microbiology | Long term monitoring of antimicrobial resistance or infection rates. | CHINET (52); Canadian Nosocomial Infection Surveillance Program (19) |
| Medicine | Departmental research | Dept. Orthopaedic Surgery, Chiba Uni. (46); Division of Hematology, Mayo Clinic (33); Division of Hematology, Jichi Medical Uni. (32) |
| Public Health | Collaborative cohort-based project to investigate public health issues (e.g., healthy eating, aging, cancer risk factors). | JACC Study (43); JPHC Study Group (33); Food4Me (27); EMAS study group (26); European Male Ageing Study (20); Global Network for Women's and Children's Health Research (19) |
| Neuro-psychology | Funded project with shared broad goal (e.g., find how "biological, psychological, and environmental factors during adolescence may influence brain development and mental health"). | IMAGEN Consortium (24) |

### 3.4.1 Nuclear physics: The ATLAS Collaboration

The ATLAS Collaboration at CERN (atlas.cern/discover/collaboration) had 755 journal articles 2010-18 in high energy or nuclear physics journals (e.g., Journal of High Energy Physics: 23%). ATLAS (A Toroidal LHC ApparatuS) at the Large Hadron Collider (LHC) is a large multinational particle detector project using specialised equipment. Its publications record particle properties, behaviours or searches (e.g., "Search for lepton-flavor violation in different-flavor, high-mass final states in $pp$ collisions at $\sqrt{s} = 13$ TeV with the ATLAS detector") including for the Higgs boson.

The role of the multinational consortium is as a fixed pool of skills to design and run the experiments. Authorship is in alphabetical order (an average of 98% alphabetical using the methods above). Although an average of 87 authors are recorded per paper, the collaboration has hundreds of authors and all are listed on all papers associated with the apparatus during their project tenure.

### 3.4.2 Astronomy: The HESS Collaboration

The HESS Collaboration (www.mpi-hd.mpg.de/hfm/HESS/pages/collaboration/) is a multinational consortium around a single stereoscopic astronomy telescope system. Its 150 publications 2004-18 record and analyse phenomena observed in space (e.g., "Discovery of very high energy γ-ray emission from Centaurus A with H.E.S.S.").

The role of the multinational consortium seems to be again a fixed pool of skills to design and run the experiments. Authorship is always in alphabetical order (an average of 95% alphabetical using the methods above), with an average of 92 recorded in Scopus but all articles having hundreds of authors.

### 3.4.3 Shared genomics task: GEBA Project

The Genomic Encyclopedia of Bacteria and Archaea (GEBA) project is a joint initiative by the Leibniz-Institute DSMZ–German Collection of Microorganisms and Cell Cultures and the U.S. Department of Energy (DOE) Joint Genome Institute (JGI) to systematically select and sequence bacterial and archaeal genomes (Wu, Hugenholtz, Mavromatis, Pukall, Dalin, et al., 2009). Its 148 Scopus publications from 2009-15 identified for the cluster had an average of 35 authors, with orders varying substantially between articles. All except one were published in the Springer environmental biology journal *Standards in Genomic Sciences*, and reported a complete, or near complete, genome for bacteria or archaea. It has sequenced 250 bacteria and archaea genomes (jgi.doe.gov/our-science/science-programs/microbial-genomics/phylogenetic-diversity), so the cluster is an incomplete record of its activities. Most of its articles had titles starting with "Complete genome sequence of" (125) or "Genome sequence of" (14). For example, one article was a short report that summarised key properties of the genome of an orange seawater bacterium, as well as reporting that the complete sequence had been reported in GenBank, with wider project results in the Genomes OnLine Database (Riedel, Held, Nolan, Lucas, Lapidus, et al., 2012).

The role of the consortium is presumably to combine the multiple types of equipment and other expertise needed to efficiently sequence and analyse genomes. Since author orders and numbers vary (although a standard list is sometimes at the end, e.g., Bristow; Eisen; Markowitz; Hugenholtz; Kyrpides; Klenk) the set seems to reflect the relative contributions of the participants. Paper authors are, on average, in 49% alphabetical order using the methods above, confirming that this is not used.

### 3.4.4  Medical cohort for a disease: Swiss HIV Cohort Study

The Swiss HIV Cohort Study (SHCS) has been from 1998 "a systematic longitudinal study enrolling HIV-infected individuals in Switzerland" (www.shcs.ch), gathered from multiple hospitals and clinics. It includes the Swiss Mother and Child HIV Cohort Study (MoCHiV). The 120 papers in the cluster were from 2000-18, had an average of 57 authors and were published in many different journals, with the most common being *AIDS* (13%) and *Journal of Infectious Diseases* (12%). The publications seem to cover a wide range of HIV-related research, and appear in many journals. There is no typical paper for the consortium. As an example, one paper reports a type of cancer risk for patients (Clifford, Rickenbach, Polesel, Dal Maso, Steffen, et al., 2008).

The role of the consortium is presumably to collect and manage information on the study. Author order varies by paper. The first few authors are typically not in alphabetical order but the remainder are (an average of 85% alphabetical using the methods above). Thus, each paper seems to have a small group of main authors: perhaps those that planned the study, conducted the analysis and wrote the paper, with the remaining authors perhaps awarded co-authorship for contributions to data collection and management.

### 3.4.5  Microbiology: Antimicrobial Resistance in China

The China Antimicrobial Surveillance Network CNINET originates from 2005 and is an ongoing surveillance program in China for antimicrobial resistance (including antibiotic resistance).   Its sister program, the China Antimicrobial Resistance Surveillance System (CARSS), focuses on regional comparisons (Hu, Zhu, Wang, & Wang, 2018) but did not appear as a consortium in the Scopus data. Only one other consortium of this type appeared in this data, from Canada, despite similar international data collection exercises (World Health Organization, 2017). CNINET's 52 Scopus publications were almost all (51) published in the Chinese Journal of Infection and Chemotherapy from 2008 to 2017 had an average of 32 authors. The publications describe resistance associated with one genus, such as, "CHINET 2009 surveillance of antibiotic resistance in enterococcus in China". This focuses on a genus of lactic acid bacteria related to human medical conditions such as urinary tract infections and meningitis. Its data was from 14 hospitals in China. There are also generic surveillance papers, such as, "CHINET 2008 surveillance of bacterial resistance in China". The value of a long term coherent consortium for this task is presumably in standardising methods for reliable comparisons over time (Hu, Guo, Zhu, Wang, Jiang, et al., 2016), as well as general efficiency gains.

The role of the consortium is presumably to manage access to data or testing at the different sites because the authors have multiple hospital affiliations. Author order varies between papers, except for the general surveillance papers. For example, the first thirteen authors were the same and in the same order for the 2008 and 2009 general surveillance papers. Authors are in an average of 50% alphabetical order (i.e., exactly that expected if alphabetical considerations are ignored) using the methods above.

### 3.4.6  Medical department research: Orthopaedic Surgery at Chiba University

The Department of Orthopaedic Surgery at Chiba University Hospital in Japan had a cluster of 46 papers from Scopus 2009-17, mainly published in Spine-related journals, such as *Spine* (37%) and *European Spine Journal* (20%), with an average of 22 authors. Its research includes both spinal and joint injuries and it seems to be highly successful award-winning research group (www.ho.chiba-u.ac.jp/en/dpt/orthopaedic.html). The main author of many papers, Professor Seiji Ohtori, had two records in Scopus that accounted for 391 publications (Author ID: 7004456445, 197 publications 2009-2014; ID: 56008400700, 194 publications 2012-2019).

The cluster is a department rather than a consortium. Most papers include only authors within the department. Authors are not in alphabetical order (average: 52%). Author order varied between papers, although often with the same first author and similar sequences of last authors and two of the three professors (as listed in the Department website in 2019) listed last except in four cases. Papers by the department not included in the cluster had fewer authors (e.g., 6: "Segmental Pedicle Screw Instrumentation and Fusion Only to L5 in the Surgical Treatment of Flaccid Neuromuscular Scoliosis"), or a smaller overlap than 80% due to a collaboration (e.g., 22 authors but several were from other universities, "Use of Bioelectrical Impedance Analysis for the Measurement of Appendicular Skeletal Muscle Mass/Whole Fat Mass and Its Relevance in Assessing Osteoporosis among Patients with Low Back Pain: A Comparative Analysis Using Dual X-ray Absorptiometry"). Thus, the cluster of 46 articles represents under a quarter of the output of the Department: a highly collaborative ad-hoc subset.

### 3.4.7   Public health risk factors: The JACC Study

The JACC Study (Japan Collaborative Cohort Study for Evaluation of Cancer Risk) included 46 papers published 2002-08, mainly in the *Journal of Epidemiology* (72%), mostly examining specific risk factors for a type of cancer (e.g., "Smoking, alcohol drinking and esophageal cancer: Findings from the JACC Study"). This study gave a lifestyle questionnaire to 73,424 people 1988-90 and tracked their health problems until the end of 1997 (Iso, Date, Yamamoto, Toyoshima, Tanabe, et al., 2002). This provided a rich data source to investigate the relationship between lifestyle factors, diseases and mortality.

The cluster is a named consortium from different universities, hospitals and health centres created for around a shared, funded task. Authors are not in alphabetic order (47% of distinct author pairs in alphabetical order). Authorship presumably reflects contributions to leading the extensive shared data gathering task for the initial questionnaires or the follow-up health outcomes information. An average of 40 authors (min: 34; max: 50) wrote each paper, with varying author order (no more than 4 papers with the same first author), although the last-listed about ten authors tended to be in a similar order. Thus, authorship seems to reflect roles in the data collection process, with the first authors presumably carrying out the analysis for a paper.

### 3.4.8   Neuropsychology: The IMAGEN Consortium

The IMAGEN Study is a consortium funded by a series of EU, national and other grants to follow a cohort of 2000 young Europeans from the age of 14 to track adolescent brain development through questionnaires, brain imaging and tests (imagen-europe.com). Its publications 2011-2016 cover different aspects of behaviour and brain development (e.g., "No differences in ventral striatum responsivity between adolescents with a positive family history of alcoholism and controls"), with an average of 26 authors per article.

The cluster contains about a quarter of the articles from the project since its website reports 104 journal articles 2010-18. Author order is typically alphabetical in the middle but not at the start (presumably the main authors for the study) and end (presumably consortium leaders). There is an average of 77% alphabetical ordering using the methods above. The role of the consortium seems to be partly to maintain and update core information about the cohort and partly to get additional information for specific analyses.

## 4   Discussion and conclusions

This paper introduces a new method to identify large publishing consortia as well as evidence of their research impact, use of alphabetical ordering and broad types. The results show that there are large publishing consortia in nuclear physics, astronomy and some health-related fields. The heuristics used to identify these consortia cannot be used to estimate the prevalence of large stable research groups because others may have different publishing agreements or may produce other types of output. Thus the 0.05% of Scopus articles included in them underestimate the extent of large publishing consortium output. The highest profile huge research collaboration in the last fifty years, the human genome project, was not included because of this. In addition, modifications to the heuristics (e.g., 70% authors in common; minimum 10 authors per paper) would produce different results. Nevertheless, the existence of publishing consortia of many types is confirmed by the examinations above of individual groups that found them all to be named and clearly identifiable entities, or at least collections of papers within departments or research groups.

Whilst the physics consortia seem to have been formed because of the expense and perhaps complexity of the equipment required for specific tasks, the health-related consortia mainly reflect the need to systematically gather large quantities of diverse high-quality human-related (life-related in the remaining cases) data to have sufficient statistical power or variety to investigate a health issue. As with the GWAS case discussed above, some health issues cannot be resolved on a small scale, or may give statistically misleading answers from repeated small-scale experiments, and so combining data is essential for progress.

The publishing consortia found have research impact that tends to be substantially, and almost universally above the world average for the fields and years in which they published. This may be due to the consortia typically involving economically advanced nations (e.g., Sud & Thelwall, 2016) or the increased likelihood to attract self-citations (Bornmann, 2017; van Raan, 1998) from the large teams but, from the example examined, it seems more likely that these teams produce research that is valuable to their disciplines, justifying the human and financial resources that seem to have been invested in many of them. Nevertheless, the results exclude all consortia that did not create at least three Scopus-indexed consortia and might (if they had no other goals) be characterised as failing and so the conclusions should not be extrapolated to non-publishing consortia.

From a scientometric perspective, to evaluate the impact of the work of individual academics or departments from their citations, it is useful to be able to attribute an appropriate share of an output to its contributors. This is complicated by publishing collaborations, especially if partial or full alphabetical ordering is used. For example, if five different groups might have produced five competing analyses of a single medical issue (e.g., whether green tea affects cancer risks) but instead publish one more powerful combined study then four fewer papers would be published and the combined paper would have many authors. Presumably economies of scale would apply (e.g., one analysis instead of five; one study design instead of five), so that the total amount of work for the combined paper would be less than five times the amount as for any individual paper. Whilst the efficiency and power improvements are good for science, a scientometric attempt to share credit for a single paper (e.g., with fractional author counting) would be unfair in this context. Conversely, for a publishing consortium that rewards participants that manage data gathering but are otherwise not active, it might be reasonable to assign fractional value (Hagen, 2013) to the less active participants. For the previously-noted biomedical phenomenon of middle author alphabetical ordering (Mongeon, Smith, Joyal, & Larivière, 2017), the middle authors (if accurately identified) could potentially be awarded equal but lesser credit to the remaining authors. Thus, a scientometric exercise would need to make a reasoned judgement about how to deal with contributions from large publishing consortia. For health-related consortia, it seems to be an ethical necessity to ensure that participation in collaborations that aid public health should not be discouraged by under-rewarding participants and so evaluators should be extremely careful when considering this option.

## 5  References

AACR Project GENIE Consortium. (2017). AACR Project GENIE: powering precision medicine through an international consortium. Cancer discovery, 7(8), 818-831.

Bernstein, B. E., Stamatoyannopoulos, J. A., Costello, J. F., Ren, B., Milosavljevic, A., Meissner, A., & Farnham, P. J. (2010). The NIH roadmap epigenomics mapping consortium. Nature Biotechnology, 28(10), 1045-1048.

Bornmann, L. (2017). Is collaboration among scientists related to the citation impact of papers because their quality increases with collaboration? An analysis based on data from F1000Prime and normalized citation scores. Journal of the Association for Information Science and Technology, 68(4), 1036-1047.

Buniello, A., MacArthur, J. A. L., Cerezo, M., Harris, L. W., Hayhurst, J., Malangone, C., & Suveges, D. (2018). The NHGRI-EBI GWAS Catalog of published genome-wide association studies, targeted arrays and summary statistics 2019. Nucleic Acids Research, 47(D1), D1005-D1012.

Clifford, G. M., Rickenbach, M., Polesel, J., Dal Maso, L., Steffen, I., Ledergerber, B., ... & Franceschi, S. (2008). Influence of HIV-related immunodeficiency on the risk of hepatocellular carcinoma. AIDS, 22(16), 2135-2141.

Costa, M. R., Qin, J., & Bratt, S. (2016). Emergence of collaboration networks around large scale data repositories: A study of the genomics community using GenBank. Scientometrics, 108(1), 21-40.

de Solla Price, D. J., (1986). Little science, big science... and beyond (p. 301). New York, NY: Columbia University Press.

Defazio, D., Lockett, A., & Wright, M. (2009). Funding incentives, collaborative dynamics and scientific productivity: Evidence from the EU framework program. Research policy, 38(2), 293-305.

Thelwall, M., & Fairclough, R. (2017). The accuracy of confidence intervals for field normalised indicators. Journal of Informetrics, 11(2), 530-540.

Fortunato, S., Bergstrom, C. T., Börner, K., Evans, J. A., Helbing, D., Milojević, S., ... & Vespignani, A. (2018). Science of science. Science, 359(6379), eaao0185 (1-7).

Garner, C. (2007). Upward bias in odds ratio estimates from genome-wide association studies. Genetic Epidemiology, 31(4), 288-295.

Gene Ontology Consortium. (2014). Gene ontology consortium: going forward. Nucleic Acids Research, 43(D1), D1049-D1056.

Gibbons, M. R., Limoges, C., Nowotny, H., Schwartzman, S., Scott, P., & Trow, M. (1994). The New Production of Knowledge: The Dynamics of Science and Research in Contemporary Societies. London, UK: Sage.

Hagen, N. T. (2013). Harmonic coauthor credit: A parsimonious quantification of the byline hierarchy. Journal of Informetrics, 7(4), 784-791.

Hoekman, J., Scherngell, T., Frenken, K., & Tijssen, R. (2013). Acquisition of European research funds and its effect on international scientific collaboration. Journal of economic geography, 13(1), 23-52.

Hu, F. P., Guo, Y., Zhu, D. M., Wang, F., Jiang, X. F., Xu, Y. C., ... & Kang, M. (2016). Resistance trends among clinical isolates in China reported from CHINET surveillance of bacterial resistance, 2005–2014. Clinical Microbiology and Infection, 22, S9-S14.

Hu, F., Zhu, D., Wang, F., & Wang, M. (2018). Current Status and Trends of Antibacterial Resistance in China. Clinical Infectious Diseases, 67(suppl_2), S128-S134.

International Human Genome Sequencing Consortium (2001). Initial sequencing and analysis of the human genome. Nature, 409(6822), 860-921.

Ioannidis, J. P. (2005). Why most published research findings are false. PLoS Medicine, 2(8), e124.

Iso, H., Date, C., Yamamoto, A., Toyoshima, H., Tanabe, N., Kikuchi, S., ... & Suzuki, H. (2002). Perceived mental stress and mortality from cardiovascular disease among Japanese men and women: the Japan Collaborative Cohort Study for Evaluation of Cancer Risk Sponsored by Monbusho (JACC Study). Circulation, 106(10), 1229-1236.

Kawashima, H., & Tomizawa, H. (2015). Accuracy evaluation of Scopus Author ID based on the largest funding database in Japan. Scientometrics, 103(3), 1061-1071.

Larivière, V., Gingras, Y., & Archambault, É. (2006). Canadian collaboration networks: A comparative analysis of the natural sciences, social sciences and the humanities. Scientometrics, 68(3), 519-533.

Larivière, V., Gingras, Y., Sugimoto, C. R., & Tsou, A. (2015). Team size matters: Collaboration and scientific impact since 1900. Journal of the Association for Information Science and Technology, 66(7), 1323-1332.

Levitt, J. M., & Thelwall, M. (2013). Alphabetization and the skewing of first authorship towards last names early in the alphabet. Journal of Informetrics, 7(3), 575-582.

Liu, X. Z., & Fang, H. (2014). The impact of publications from mainland China on the trends in alphabetical authorship. Scientometrics, 99(3), 865-879.

Moed, H. F., & Plume, A. (2013). Studying scientific migration in Scopus. Scientometrics, 94(3), 929-942.

Mongeon, P., Smith, E., Joyal, B., & Larivière, V. (2017). The rise of the middle author: Investigating collaboration and division of labor in biomedical research using partial alphabetical authorship. PloS ONE, 12(9), e0184601.

Moore, F. A., Moore, E. E., Billiar, T. R., Vodovotz, Y., Banerjee, A., & Moldawer, L. L. (2017). The role of NIGMS P50 sponsored team science in our understanding of multiple organ failure. The journal of trauma and acute care surgery, 83(3), 520-531.

Munafò, M. R., Nosek, B. A., Bishop, D. V., Button, K. S., Chambers, C. D., Du Sert, N. P., & Ioannidis, J. P. (2017). A manifesto for reproducible science. Nature Human Behaviour, 1(1), 0021.

Olds, J. L. (2016). The rise of team neuroscience. Nature Reviews Neuroscience, 17(10), 601.

Pe'er, I., Yelensky, R., Altshuler, D., & Daly, M. J. (2008). Estimation of the multiple testing burden for genomewide association studies of nearly all common variants. Genetic Epidemiology, 32(4), 381-385.

Psaty, B. M., O'donnell, C. J., Gudnason, V., Lunetta, K. L., Folsom, A. R., Rotter, J. I., & Boerwinkle, E. (2009). Cohorts for Heart and Aging Research in Genomic Epidemiology (CHARGE) Consortium: Design of prospective meta-analyses of genome-wide association studies from 5 cohorts. Circulation: Cardiovascular Genetics, 2(1), 73-80.

Riedel, T., Held, B., Nolan, M., Lucas, S., Lapidus, A., Tice, H., ... & Goodwin, L. A. (2012). Genome sequence of the orange-pigmented seawater bacterium Owenweeksia hongkongensis type strain (UST20020801 T). Standards in Genomic Sciences, 7(1), 120.

Roberts, L. (2001). Timeline: A history of the Human Genome Project. Science, 291(5507), 1195-1200.

Sjögårde, P., & Ahlgren, P. (2018). Granularity of algorithmically constructed publication-level classifications of research publications: Identification of topics. Journal of Informetrics, 12(1), 133-152.

Sud, P., & Thelwall, M. (2016). Not all international collaboration is beneficial: The Mendeley readership and citation impact of biochemical research collaboration. Journal of the Association for Information Science and Technology, 67(8), 1849-1857.

Thelwall, M. (2017). Three practical field normalised alternative indicator formulae for research evaluation. Journal of Informetrics, 11(1), 128-151.

van Raan, A. (1998). The influence of international collaboration on the impact of research results: Some simple mathematical considerations concerning the role of self-citations. Scientometrics, 42(3), 423-428.

Vermeulen, N., Parker, J. N., & Penders, B. (2013). Understanding life together: A brief history of collaboration in biology. Endeavour, 37(3), 162-171.

Wagner, C. S., Whetsell, T. A., & Mukherjee, S. (in press). International research collaboration: Novelty, conventionality, and atypicality in knowledge recombination. Research Policy.

Waltman, L., Van Eck, N. J., & Noyons, E. C. (2010). A unified approach to mapping and clustering of bibliometric networks. Journal of Informetrics, 4(4), 629-635.

Waltman, L. (2012). An empirical analysis of the use of alphabetical authorship in scientific publishing. Journal of Informetrics, 6(4), 700-711.

Welter, D., MacArthur, J., Morales, J., Burdett, T., Hall, P., Junkins, H., & Parkinson, H. (2013). The NHGRI GWAS Catalog, a curated resource of SNP-trait associations. Nucleic Acids Research, 42(D1), D1001-D1006.

World Health Organization. (2017). Global antimicrobial resistance surveillance system (GLASS) report: early implementation 2016-2017. https://www.who.int/glass/resources/publications/early-implementation-report/en/

Wu, D., Hugenholtz, P., Mavromatis, K., Pukall, R., Dalin, E., Ivanova, N. N., & Hooper, S. D. (2009). A phylogeny-driven genomic encyclopaedia of Bacteria and Archaea. Nature, 462(7276), 1056-1060.

Wuchty, S., Jones, B. F., & Uzzi, B. (2007). The increasing dominance of teams in production of knowledge. Science, 316(5827), 1036-1039.

Xie, Z., Ouyang, Z., & Li, J. (2016). A geometric graph model for coauthorship networks. Journal of Informetrics, 10(1), 299-311.